\documentclass[aps,prb,twocolumn,floatfix,showpacs]{revtex4}
\usepackage{amsmath,amssymb,graphicx,bm}
\begin{document}
\title{Thermodynamic origin of order parameters in mean-field models of
  spin glasses}

\author{V.  Jani\v{s}} \author{ L. Zdeborov\'a}

\affiliation{Institute of Physics, Academy of Sciences of the Czech
  Republic, Na Slovance 2, CZ-18221 Praha 8, Czech Republic}
\email{janis@fzu.cz, zdeborl@fzu.cz}

\date{\today}


\begin{abstract}
  We analyze thermodynamic behavior of general $n$-component mean-field
  spin glass models in order to identify origin of the hierarchical
  structure of the order parameters from the replica-symmetry breaking
  solution. We derive a configurationally dependent free energy with local
  magnetizations and averaged local susceptibilities as order parameters.
  On an example of the replicated Ising spin glass we demonstrate that the
  hierarchy of order parameters in mean-field models results from the
  structure of inter-replica susceptibilities. These susceptibilities serve
  for lifting the degeneracy due to the existence of many metastable states
  and for recovering thermodynamic homogeneity of the free energy.
\end{abstract}

\maketitle 
\section{Introduction}\label{sec:Intro}

An effective spin exchange between magnetic impurities diluted in a host
metallic matrix is inhomogeneous and does not prefer either ferromagnetic
or antiferromagnetic alignment of the impurity spins. It is hence natural
to simulate such a magnetic behavior by a random distribution of the spin
exchange as introduced by Edwards and Anderson.\cite{Edwards75} Frustration
in the spin exchange induced by static randomness has since then become the
hallmark of microscopic models of spin glasses. Due to complexity of
spin-glass models, majority of theoretical research of spin glasses has
concentrated on mean-field properties of the Ising spin glass proposed by
Sherrington and Kirkpatrick.\cite{Sherrington75,Binder86} Although we have
by now a consistent comprehensive solution of the Sherrington-Kirkpatrick
(SK) model in form of the replica-symmetry breaking (RSB) scheme of
Parisi,\cite{Parisi80} there remain a few issues where a conclusive answer
is still pending.
 
One of unsettling features in spin-glass models is the way the mean-field
solution and its interpretation are obtained. On one hand, the replica
trick and replicas of the spin variables are used to convert the static
randomness to a dynamical, thermal-like averaging and to offer a necessary
space for introduction of symmetry breaking order parameters.  They are
overlaps between different spin replicas $Q^{ab} = N^{-1}\sum_i\langle
S_i^a S_i^b\rangle$. The brackets denote joint thermal and configurational
averaging. The physical interpretation of the replica order parameters can,
however, be reached only within a thermodynamic approach where a mean-field
free energy is first constructed for typical configurations of the
spin-exchange. In this thermodynamic approach pioneered by Thouless,
Anderson and Palmer (TAP)\cite{Thouless77} we have only local
magnetizations $m_i = \langle S_i\rangle_T$, thermally averaged local
spins, as order parameters for the Ising spin glass. To complete the
mean-field solution, a relation between the local magnetizations from the
TAP approach and the averaged replica parameters must be established. It
was successfully done within the so-called cavity method with many
metastable TAP states.\cite{Mezard86,Mezard87}

The existence of a multiple of solutions to the TAP equations has been
demonstrated both directly\cite{Bray79,Cavagna04} and also
indirectly.\cite{Bray80,Cavagna03} Within the cavity method the parameters
$q^{\alpha\beta} = N^{-1}\sum_i m_i^\alpha m_i^\beta$, overlaps between
local magnetizations from different pure (metastable) states, were related
to the RSB parameters $Q^{ab}$. However, the order parameters from the RSB
construction can generally be represented as $Q^{ab} = \chi^{ab} + q^{ab}$
where $q^{ab} = N^{-1}\sum_i m_i^a m_i^b$ is the overlap of local
magnetizations between different spin replicas (should not be identified
with different metastable states) and $\chi^{ab} = N^{-1}\sum_i (\langle
S_i^a S_i^b\rangle_T - m_i^a m_i^b)$ is the local overlap susceptibility.
It is just only the special case of the non-replicated Ising model
($S_i^2=1$) where the overlap susceptibility in the TAP construction with a
single equilibrium state can be expressed via local magnetizations. If we
investigate vector spin glasses\cite{Fischer91} or assume the existence of
many quasi-equilibrium states we cannot get rid of local susceptibilities
as independent order parameters.

It indeed appears that the local magnetizations $m_i$ of the TAP theory
fail in determining unambiguously the equilibrium thermodynamic states. It
has been demonstrated by numerical means that there is a macroscopic
portion of configurations of the spin exchange for which we are either
unable to find a stable solution of the TAP equations at low
temperatures\cite{Bray79,Nemoto85} or we have paired stable and unstable
solutions with nearly the same energies.\cite{Cavagna04} It means that
macroscopic parameters, temperature and external magnetic field, do not
determine uniquely local magnetizations and it is unclear whether there is
an equilibrium TAP state at all temperatures. In such a case the TAP free
energy is no longer thermodynamically homogeneous and solutions of the TAP
equations depend on initial/boundary conditions. We have to introduce spin
replicas and overlap local susceptibilities emerge as natural order
parameters.\cite{Janis03}
 
The aim of this paper is to trace down the genesis of the order parameters
$Q^{ab}$ from the RSB scheme within the thermodynamic TAP-like approach.
In particular, we analyze the role of local magnetizations $m_i^a$ with
their averaged overlaps $q^{ab}$ and averaged local susceptibilities
$\chi^{ab}$ in the mean-field free energy of a general $n$-component
spin-glass model with spin components $S^a$, $a=1,\ldots,n$. We employ the
replicated TAP theory from Ref.~\onlinecite{Janis03} as the simplest
example of an $n$-component spin model and find that not the local
magnetizations and their averaged overlaps $q^{ab}$, but rather the overlap
local susceptibilities $\chi^{ab}$ seem to be relevant order parameters in
the spin glass phase. To demonstrate this we first analyze two real
replicas of Ising spin variables. Using the replica-symmetric ansatz we
then continue analytically the replicated TAP free energy to arbitrary
replication factors to enable investigation of thermodynamic homogeneity.
We recover the one-step RSB solution of Parisi by minimizing thermodynamic
inhomogeneity incurred by the imposed replica-symmetry.  Conditions for
global thermodynamic homogeneity and a way to reach a thermodynamically
homogeneous mean-field free energy are presented.

The paper is organized as follows. We define in Sec.~\ref{sec:Models} the
studied models being generally $n$-component spin systems. Multi-component
spin models can arise either due to the vector character of spins or due to
replications of the phase space introduced to test thermodynamic
homogeneity. We sum in Sec~\ref{sec:Multiple} mean-field thermal
fluctuations for a typical configuration of the spin exchange before
averaging over randomness. In Sec.~\ref{sec:Finite} we analyze the case of
two real replicas to demonstrate the importance of the averaged local
overlap susceptibility in the spin-glass phase. We analytically continue
the replicated free energy from integer numbers of real replicas to
arbitrary replication factors in Section~\ref{sec:1RSB} using the
replica-symmetric ansatz. Stability conditions and a hierarchical
construction of a globally homogeneous solution are presented in
Sec.~\ref{sec:Stability}. Finally, in Sec.~\ref{sec:Interpretation} we
discuss the physical meaning of the order parameters in the replicated
phase space and we summarize the conclusions in Sec.~\ref{sec:Conclusions}.

\section{Spin-glass models and thermodynamic homogeneity}
\label{sec:Models}

The Ising spin glass in its mean-field limit, the SK model, is degenerate
in that the fluctuation-dissipation theorem allows us to exclude local
susceptibilities from the order parameters of the configurationally
dependent free energy. We hence will consider general vector spin models so
as to assess better the role and importance of local susceptibilities in
the mean-field thermodynamics of spin glasses.  Our starting Hamiltonian
reads
 \begin{equation}\label{eq:Hamiltonian}
 \widehat{H} = -\frac 12 \sum_{i\neq j} J_{ij} \vec{S}_i\cdot\vec{S}_j -
\vec{H} \cdot \sum_i \vec{S}_i\ .
 \end{equation}
 The norm of the spin vectors is fixed and we assume for our purposes that
 it depends on the number of spin components $n$, that is
\begin{equation}\label{eq:Spin-norm}
\vec{S}_i\cdot\vec{S}_i = \sum_{a=1}^n S_i^aS_i^a
= n s_n^2 \ .
\end{equation}
We are interested in the thermodynamic limit of thermodynamic potentials of
this Hamiltonian for fixed configurations of the spin-exchange parameters
$J_{ij}$. The free energy reads
\begin{multline}\label{eq:Free-energy}
  F = - \frac 1\beta \lim_{N\to\infty}\ln\text{Tr}_S\left[ \exp\left\{
      \frac \beta 2 \sum_{i,j=1}^N J_{ij} \vec{S}_i\cdot\vec{S}_j
    \right.\right.  \\ \left. \left. + \beta \vec{H}\cdot \sum_{i=1}^N
      \vec{S}_i\right\} \right]\ .
\end{multline}
The trace $\text{Tr}_S$ is taken over all admissible spin configuration
respecting the normalization condition~\eqref{eq:Spin-norm}.  The free
energy in Eq.~\eqref{eq:Free-energy} is very general and covers vector spin
models as well as replicated spin models that we will need for the
investigation of thermodynamic homogeneity.
 
Since the ordering part of the spin exchange is irrelevant for our
reasoning we assume only purely fluctuating frustrated spin exchange that
in the mean-field limit has a long-range character and is a Gaussian random
variable with
\begin{equation}\label{eq:Exchange-norm}
\left\langle J_{ij}\right\rangle_{av} = 0,\qquad \left\langle
J_{ij}^2\right\rangle_{av} = \frac {J^2}N,\qquad \text{for}\quad  i\neq j.
\end{equation}

The mean-field models are peculiar in that they are long-range with a
volume-dependent spin-exchange as in Eq.~\eqref{eq:Exchange-norm}.  The
volume-dependence just compensates for the infinite range of the spin
exchange. Thereby the energy remains linearly proportional to the volume
and we can expect that the mean-field solution emulates the thermodynamic
properties of realistic models in a sense reliably or at least
consistently.

One of fundamental properties of thermodynamic systems is thermodynamic
homogeneity. It says that thermodynamic potentials are products of the
volume and functions where all extensive variables enter only via their
spatial densities. Thermodynamic homogeneity is normally expressed as the
Euler lemma
\begin{multline}\label{eq:Euler}
  \alpha F(T,V,N,\ldots,X_i,\ldots) \\
  = F(T, \alpha V, \alpha N,\ldots,\alpha X_i,\ldots)
\end{multline}
where $\alpha$ is an arbitrary positive number and the set $\{X_i\}$ covers
all extensive variables, the thermodynamic potential $F$ depends on. Only
if this homogeneity condition is fulfilled by thermodynamic potentials we
can prove the Gibbs-Duhem relation and the thermodynamic limit does not
depend on the shape of the volume and on boundary/initial conditions. The
simplest example of a thermodynamically inhomogeneous systems is the
classical ideal gas with distinguishable particles (Gibbs paradox).
 
Thermodynamic homogeneity is a consequence of invariance of the
thermodynamic limit of short-range models with respect to scalings
(contractions and dilatations) of the phase space. Since the spin-exchange
in mean-field models is an extensive variable, it is not \'a priori evident
whether also mean-field models should be thermodynamically homogeneous. The
mean-field thermodynamics is well defined if the thermodynamic limit
exists. Mean-field thermodynamic potentials must be thermodynamically
homogeneous to make the thermodynamic limit meaningful. Scalings in the
phase space of mean-field models are nonlinear transformations due to the
volume dependence of the spin exchange, therefore they are replaced by
replications of the phase-space variables when testing thermodynamic
homogeneity. To introduce real replicas we use an identity for
integer~$\nu$: $\left[\text{Tr}\ e^{-\beta H}\right]^\nu = \text{Tr}_\nu
\exp\left\{\beta\sum\limits_{a=1}^{\nu}H^\alpha\right\} = \text{Tr}_\nu
\exp\left\{\beta\sum\limits_{a=1}^{\nu} \left(\sum_{i,j} J_{ij}S_i^a S_j^a
    + \sum_i S_i^a\right)\right\}$, where each replicated spin variable
$S_i^a$ is treated independently, i.~e., the trace operator $\text{Tr}_\nu$
operates on the $\nu$-times replicated phase space. The free energy of a
$\nu$-times replicated system must be just $\nu$-times the free energy of
the non-replicated one. To test robustness of this property we add a small
perturbation breaking the replica independence $\Delta H(\mu)= \sum_i
\sum_{a < b} \mu^{ab} S_i^a S_i^b$ to allow replica-symmetry breaking in
the system response to this disturbance.  The perturbed free energy per
replica reads
\begin{multline}\label{eq:avFE}
  F_\nu (\mu) = - \frac 1\beta\ \frac 1\nu \ln\text{Tr}_S\left[
    \exp\left\{-\beta\sum_{\alpha=1}^\nu H^\alpha \right.\right. \\
  \left.\left.  -\beta \Delta H(\mu) \right.\bigg\}\right.\bigg]\ .
\end{multline}
If the system is thermodynamically homogeneous and the equilibrium state
stable, the result must be independent of the number of real replicas after
switching off the perturbation, that is
\begin{equation}\label{eq:av-homogeneity}
\frac{d}{d\nu}\lim_{\mu\to0}F_\nu(\mu) \equiv 0 \ .
\end{equation}
This is a quantification of thermodynamic homogeneity of long-range,
mean-field models.  If this condition is fulfilled the thermodynamic limit
and macroscopic quantities are well defined. This is the eventual situation
we would like to achieve also in spin-glass models.
Equation~\eqref{eq:av-homogeneity} must hold for the configurationally
dependent as well as for the averaged free energy.

Thermodynamic homogeneity expresses invariance of the thermodynamic limit,
that is, invariance of the way the number of lattice sites becomes
infinite. Both limits $N\to\infty$ and $N'=\nu N\to\infty$ generate the
same macroscopic equilibrium state in thermodynamically homogeneous
systems.  Thermodynamic homogeneity can be formulated also in the replica
trick, where is is equivalent to invariance of the limit of the number of
replicas to zero.  If the system is thermodynamically homogeneous then the
limits $n\to0$ and $\nu n\to0$ must produce the same results. We hence can
introduce real replicas also in the replica trick to test thermodynamic
homogeneity there.\cite{Janis04}

\section{Mean-field theory for $n$-component spin glass models}
\label{sec:Multiple}

Models with $n$ independent spin components have the advantage that they
can be treated in the same way as their replicated versions. The only
difference is in the number of spin components and the way we perform the
trace over spins.  We hence will not specify the trace operator in this
section and keep the result valid for all situations of interest.

A mean-field theory is an approximation where the interacting part of the
free energy is replaced by purely local terms. It can be achieved either by
a volume-dependent scaling of a long-range exchange as in the SK model or
we can consider the limit to infinite dimensions with appropriately
rescaled short-range spin exchange in the Edwards-Anderson model.  The best
way to control a mean-field approximation is to employ a perturbation
expansion in the spin exchange $J_{ij}$. The mean-field contribution is
then the leading nonvanishing order in the long-range or
infinite-dimensional limit.  It is easy to verify that spins on distinct
lattice sites interact in the mean-field solution of spin-glass models with
the normalization of the spin exchange from Eq.~\eqref{eq:Exchange-norm}
only if connected by maximally squares of $J_{ij}$. This property of the
mean-field solution can be conveniently represented diagrammatically. Then
only trees and simple loops in the spin exchange contribute.  Trees
represent an inhomogeneous Weiss mean-field while the loops stand for
Onsager's cavity field.

\subsection{Tree contribution}
\label{sec:Tree}

Tree contributions in the diagrammatic expansion for the free energy lead
to local magnetizations as thermally averaged values of spin variables. We
write
\begin{equation}\label{eq:Spin-fluctuation}
\vec{S}_i = \vec{M}_i + \delta\vec{S}_i
\end{equation}
where $\vec{M}_i = \text{Tr}_S[\vec{S}_i \widehat{\rho}]$, $\widehat{\rho}
= \exp\{-\beta\widehat{H}\}/Z$, $\text{Tr}_S[\widehat{\rho}]=1$,. is the
local magnetization and $\delta \vec{S}_i$ is the fluctuating part of the
spin vector. We rewrite the free energy as a contribution from local
magnetizations and fluctuating spins
\begin{multline}\label{eq:FE-tree}
  F = -\frac 12 \sum_{ij} J_{ij}\vec{M}_i\cdot\vec{M}_j - \vec{H}\cdot
  \sum_i \vec{M}_i \\ - \frac 1\beta \ln \text{Tr}_{\delta S}\left[
    \exp\left\{\frac \beta 2 \sum_{ij} J_{ij} \delta\vec{S}_i \cdot
      \delta\vec{S}_j \right.  \right. \\ \left.\left.  + \beta \sum_i
      \delta\vec{S}_i \cdot \left(\vec{H} + \sum_j J_{ij}
        \vec{M}_j\right)\right\}\right]
\end{multline}
where the trace is taken only over spin fluctuations around their thermal
averages. Hence, $\text{Tr}_{\delta S}[\delta \vec{S}_i\widehat{\rho}]=0$.
The Weiss mean-field solution is obtained from Eq.~\eqref{eq:FE-tree} if
the quadratic term in spin fluctuations is neglected. Since
$J_{ij}^2\propto N^{-1}$ we cannot neglect correlations between fluctuating
spin components in spin-glass models.

\subsection{One-loop contribution}
\label{sec:Loop}

To derive the leading loop corrections to the inhomogeneous Weiss mean
field we reformulate the problem in Gaussian fluctuating fields. We can
represent free energy \eqref{eq:FE-tree} with Gaussian fluctuating fields
as follows
\begin{multline}\label{eq:FE-Gauss-represent}
  F = -\frac 12 \sum_{ij} J_{ij}\vec{M}_i\cdot\vec{M}_j - \vec{H}\cdot
  \sum_i \vec{M}_i + \frac 1\beta \ln Z_0 \\ - \frac 1\beta \ln \left[ \int
    D\Phi \exp\left\{- \frac 12 \sum_{ij} \vec{\Phi}_i \cdot
      W_{ij}\vec{\Phi}_j \right\} \right. \\ \left. \text{Tr}_{\delta S}
    \exp\left\{ \sum_i \delta\vec{S}_i \cdot \left(\vec{\Phi}_i +
        \beta\vec{H} + \sum_j \beta J_{ij} \vec{M}_j\right)\right\}\right]
\end{multline}
where $Z_0 = \int D\Phi \exp\left\{- \frac 12 \sum_{ij} \vec{\Phi}_i \cdot
  W_{ij}\vec{\Phi}_j \right\}$, $\int D\Phi =
\prod_i\prod_a\int_{-\infty}^\infty d\Phi_i^a/\sqrt{2\pi}$, and
$\widehat{W} = \widehat{J}^{-1}$. We use hats for operators (matrices) in
the lattice space. Notice that the spin-exchange matrix $\widehat{J}$ only
with off-diagonal elements is not invertible. We, however, can introduce a
suitable chemical potential $- \kappa\delta_{ij}$ to the spin exchange so
that $\widehat{J}_\kappa = -\widehat{\kappa} + \widehat{J}$ is invertible.
It will become evident from the construction that the result at the end
will be independent of this artificial chemical potential.

Free energy \eqref{eq:FE-Gauss-represent} is difficult in that the trace
over the spin fluctuations makes an effective potential for the Gaussian
fields $\vec{\Phi}_i$ complicated. The potential is, however, local. The
one-loop corrections to the Gaussian model can be incorporated in a local
self-energy. This self-energy is generally a matrix in the spin components
and we denote it $\underline{\sigma}_i$. We underline matrices in spin
indices. The local self-energy is calculated in a way similar to the cavity
method. The procedure is called a generalized coherent potential and was
used to derive mean-field theories of quantum itinerant
models.\cite{Janis89,Georges96}The one-loop free energy is that of a
Gaussian model with an inhomogeneous, site-diagonal potential
$\widehat{\underline{\sigma}}$ into which we embed coherently impurities
with the medium potential $\underline{\sigma}_i$ replaced by the actual
local interaction of the random fields from
Eq.~\eqref{eq:FE-Gauss-represent}. It is important that the embedding of
impurities is coherent. That is, the fluctuating fields are correlated via
local elements of a renormalized exchange of the medium and the effect of
the potential $\underline{\sigma}_i$ is locally the same as that of the
active interaction. It means that if we remove a site $i$ from the Gaussian
model with a local potential $\underline{\sigma}_i$ and replace it with the
local interaction between the fluctuating fields from the original free
energy, there is no macroscopic change to be observed.

To perform this task quantitatively we first renormalize the nonlocal
correlation matrix $\widehat{W}\to \widehat{W} -\underline{\widehat{\sigma}
  }$.  Although the bare propagator $\widehat{W}$ is diagonal in spin
components, the self-energy, as an interaction-induced response to the
symmetry-breaking term $\Delta H(\mu)$ can become a nontrivial matrix in
spin indices even after switching off the perturbation $\mu^{ab}\to 0$.  We
introduce local response matrices $\underline{C}_{ii} =\left [(\widehat{W}
  -\underline{\widehat{\sigma}})^{-1}\right]_{ii}$ containing the
correlation between the local sites with fluctuating fields and the
surrounding medium of the Gaussian model. Since we have to remove the
potential $\underline{\sigma}_i$ at the active sites, the effective
exchange between the active locally fluctuating fields will be
$\widehat{W}\to\underline{ C}_{ii}^{-1} + \underline{\sigma}_i$ where the
inversion is restricted to the chosen lattice site. The one-loop free
energy then reads
\begin{multline}\label{eq:FE-1L1}
  F = -\frac 12 \sum_{ij} J_{ij}\vec{M}_i\cdot\vec{M}_j - \vec{H}\cdot
  \sum_i \vec{M}_i \\ + \text{Tr}\ln(1 -
  \underline{\widehat{\sigma}}\beta\widehat{J}) - \frac 1\beta \sum_i \ln
  \int D\Phi \\ \times\exp\left\{- \frac 12 \vec{\Phi} \cdot
    \underline{C}_{ii} [ 1 +
    \underline{\sigma}_i\underline{C}_{ii}]^{-1}\Phi\right\} \\
  \text{Tr}_{\delta S} \exp\left\{ \delta\vec{S}_i \cdot \left(\vec{\Phi} +
      \beta\vec{H} + \sum_j \beta J_{ij} \vec{M}_j\right)\right\}\ .
\end{multline}
All introduced local thermodynamic variables $\vec{M}_i$,
$\underline{\sigma}_i$, and $\underline{C}_{ii}$ enter the free energy as
variational parameters and are determined from the respective stationarity
equations.\cite{Janis89}

We can now perform the integration over the fluctuating Gaussian fields
$\vec{\Phi}_i$ and return to the trace with full spins. We obtain
\begin{multline}\label{eq:FE-1L2}
  F = \frac 12 \sum_{ij} J_{ij}\vec{M}_i\cdot\vec{M}_j - \frac 1{2\beta}
  \sum_i \vec{M}_i\cdot [\underline{C}_{ii}^{-1} +
  \underline{\sigma}_i]^{-1}\vec{M}_i \\ + \frac 1{2\beta}\text{Tr}\ln(1 -
  \underline{\widehat{\sigma}}\beta \widehat{J}) + \frac 1{2\beta} \sum_i
  \text{Tr}\ln[1 + \underline{\sigma}_i\underline{C}_{ii}]\\ -\frac 1\beta
  \sum_i \ln\left[ \text{Tr}_{S} \exp \left\{\frac 12 \vec{S}_i \cdot [
      \underline{C}_{ii}^{-1} +
      \underline{\sigma}_i]^{-1} \vec{S}_i\right.\right. \\
  \left.\left.  \vec{S}_i \cdot\left( \beta\vec{H} + \sum_j \beta J_{ij}
        \vec{M}_j - [ \underline{C}_{ii}^{-1} +
        \underline{\sigma}_i]^{-1}\vec{M}_i \right )\right\}\right]\ .
\end{multline}
We can interpret the individual terms in the mean-field free
energy~\eqref{eq:FE-1L2} straightforwardly. The first term is the energy
due to local magnetizations. The second term is a self-interaction of local
magnetizations due to Onsager's cavity filed. The third term is the free
energy of a Gaussian model with a potential $\widehat{\underline{\sigma}}$.
The fourth term is a subtraction of local sites from the Gaussian medium on
which the spins interact with both the Weiss and Onsager's fields contained
in the last term.

Free energy~\eqref{eq:FE-1L2} is an approximate solution where all tree and
one-loop diagrams have been summed. Such an approximation becomes exact in
the long-range limit with condition~\eqref{eq:Exchange-norm} or in infinite
dimensions with $\langle J_{ij}^2\rangle_{av} = J^2/2d$.\cite{Janis89} In
this limit we can further simplify this representation, in particular the
nonlocal contribution from the Gaussian model. We do it diagrammatically
(expansion in the spin exchange). Instead of summing the contributions from
the spin exchange in the logarithm we find the mean-field form of the local
correlation function. We have
\begin{equation}\label{eq:Trace-log}
\underline{C}_{ii} = - \frac \delta{\delta\underline{\sigma}_i} \ln (1 -
\underline{\widehat{\sigma}}\beta \widehat{J}) = \left[\beta \widehat{J}(1
- \underline{\widehat{\sigma}}\beta\widehat{J})^{-1}\right]_{ii}\ .
\end{equation}
Once we find an appropriate representation for the diagonal elements of the
correlation function $\underline{C}_{ii}$ we integrate it back to obtain a
representation of the logarithm.

We know that in the mean-field spin-glass model with
Eq.~\eqref{eq:Exchange-norm} distinct lattice sites are connected by just
two bonds $J_{ij}^2$. We then have to create couples of spin exchanges in
the expansion of $\underline{C}_{ii}$ to which we ascribe identical lattice
coordinates. When lattice sites $i,j$ are interconnected by $J_{ij}^2$
there is no further correlation between these sites. If we fix one site
index, the other one contributes only as an average over the lattice. Not
to go beyond the one-loop approximation, contractions connecting spin
couplings with identical lattice coordinates must not cross and only
contractions within contractions can take place (non-crossing
approximation). The self energy renormalized by all single loop
contributions becomes a new vertex function. We can represent this
renormalization self-consistently as
\begin{equation}\label{eq:chi-def}
\underline{\chi}_{ii} = \underline{\sigma}_i \left[1 - \underline{\sigma}_i
\sum_j \beta^2J_{ij}^2 \underline{\chi}_{jj}\right]^{-1}\ .
\end{equation}
With this result the correlation function $\underline{C}_{ii}$ reads
\begin{equation}\label{eq:Cii}
\underline{C}_{ii} = \sum_j \beta^2 J_{ij}^2 \underline{\chi}_{jj} \left[1
- \underline{\sigma}_i \sum_j \beta^2J_{ij}^2
\underline{\chi}_{jj}\right]^{-1}\ .
\end{equation}
Using Eqs.~\eqref{eq:chi-def} and~\eqref{eq:Cii} we obtain the mean-field
limit of the nonlocal logarithm from Eq.~\eqref{eq:FE-1L2}
\begin{multline} \label{eq:log-repr}
  \text{Tr} \ln (1 - \underline{\widehat{\sigma}}\beta \widehat{J}) = \frac
  12 \sum_{ij} \beta^2J_{ij}^2 \text{Tr}[\underline{\chi}_{ii}
  \underline{\chi}_{jj} ] \\ + \sum_i \text{Tr}\ln\left(1 -
    \underline{\sigma}_i \sum_j \beta^2J_{ij}^2
    \underline{\chi}_{jj}\right)
\end{multline}
where the first term on the right-hand side is a compensation for the
$\underline{\sigma}$-dependence of $\underline{\chi}$.  We can replace the
nonlocal logarithm in Eq.~\eqref{eq:FE-1L2} by the right-hand side of
Eq.~\eqref{eq:log-repr}. We then have to add the new local vertex functions
$\underline{\chi}_{ii}$ to the variational parameters and obtain a
mean-field free energy for multi-component spin-glass models.

The phase space with thermodynamic parameters $\vec{M}_i$,
$\underline{\sigma}_i$, $\underline{C}_{ii}$, and $\underline{\chi}_{ii}$
is unnecessarily large. We can explicitly utilize the solutions of
stationarity equations for $\underline{\sigma}_i$ and $\underline{C}_{ii}$
so that all loop quantities are expressed only via the vertex functions
$\underline{\chi}_{ii}$. These equations are just Eqs.~\eqref{eq:chi-def}
and~\eqref{eq:Cii}. Moreover, we realize that the vertex functions
$\underline{\chi}_{ii}$ appear in the free energy only in their averaged
form and we can introduce new configurationally averaged loop parameters
$\beta^2J^2\underline{\chi} = \sum_j\beta^2 J_{ij}^2
\underline{\chi}_{jj}$. Using all these simplifications in free
energy~\eqref{eq:FE-1L2} we finally end up with a generalized TAP free
energy for $n$-component spin-glass models
\begin{multline}\label{eq:FE-final}
  F= \frac 12 \sum_{ij} J_{ij}\vec{M}_i\cdot\vec{M}_j - \frac{\beta J^2}2
  \sum_i \vec{M}_i\cdot\underline{\chi}\vec{M}_i \\ + \frac {\beta J^2}4 N
  \text{Tr}[\underline{\chi}\underline{\chi}] - \frac 1\beta \sum_i
  \ln\left[ \text{Tr}_S \exp\left\{ \frac {\beta^2J^2}2
      \vec{S}_i\cdot\underline{\chi}\vec{S}_i\right.\right. \\ \left.\left.
      +\ \beta\vec{S}_i\cdot\left(\vec{H} + \sum_j J_{ij}\vec{M}_j - \beta
        J^2 \underline{\chi}\vec{M}_i\right)\right\}\right]\ .
\end{multline}
This mean-field free energy is very general and covers the vector spin
models as well as replicated models used for investigation of thermodynamic
homogeneity.

It is straightforward to find that the saddle-point equation for the
averaged vertex reads
\begin{equation}\label{eq:chi-eq} \underline{\chi} = \frac 1N \sum_j
\left(\left\langle \vec{S}_j\vec{S}_j\right\rangle - \vec{M}_j
\vec{M}_j\right)
\end{equation}
where the vectors on the right-hand side form a tensor (matrix) in spin
components.  From this equation we obtain the fluctuation-dissipation
theorem for vector spin models
\begin{equation}\label{eq:chi-diag}
\text{Tr}\ \underline{\chi} = \frac 1N\sum_j (n s_n^2 -
\vec{M}_j\cdot\vec{M}_j)\ .
\end{equation}
We see from Eq.~\eqref{eq:chi-eq} that the averaged vertex
$\underline{\chi}$ is nothing but an averaged local susceptibility (up to
the factor~$\beta$). Only in the isotropic case with vanishing off-diagonal
elements of $\underline{\chi}$ we get rid of the local susceptibility as an
order parameter in the TAP-like free energy. We, however, know that this is
not the case either in vector models\cite{Gabay81} or in the replicated
Ising spin glass.\cite{Janis03}

\subsection{Configurational averaging}
\label{sec:Averaging}

Free energy~\eqref{eq:FE-final} has not yet the form of a typical
mean-field representation, being standardly a single-site theory. To
convert the spatially inhomogeneous free energy to a single-site expression
we have to average over the spin-exchange configurations. It cannot,
however, be done without a few assumptions.  First of all, we have to
assume that the local magnetizations $\vec{M}_i$ and averaged
susceptibilities $\underline{\chi}$ describe uniquely thermodynamic
equilibrium states. It means that we can find a set of these parameters
with which the free energy is minimal. It is the case if the nonlocal
susceptibility is nonnegative. That is, the second derivative
\begin{equation}\label{eq:Susceptibility-nonlocal}
\frac{\delta^2 \beta F}{\delta M_i^a \delta M_j^b} = -\beta
J_{ij}\delta^{ab} +
\delta_{ij}\left(\left[\underline{\chi}_{ii}^{-1}\right]^{ab} +\beta^2 J^2
\chi^{ab}\right)
\end{equation}
has no negative eigenvalues. We denoted $\underline{\chi}_{ii}$ the local
susceptibility at site $i$. It is a matrix defined as in
Eq.~\eqref{eq:chi-eq} but without averaging over the lattice sites.  It is
a straightforward task to find that the nonlocal susceptibility, inverse of
the matrix from Eq.~\eqref{eq:Susceptibility-nonlocal}, has nonnegative
eigenvalues if a generalized de Almeida-Thouless (AT) condition is
fulfilled
\begin{equation}\label{eq:AT-generalized}
\text{det}\left(\underline{1} - \frac{\beta^2
J^2}N \sum_i \underline{\chi}_{ii}^2\right) \ge 0\ .
\end{equation}
This inequality in fact stands for $n$ conditions $1\ge \beta^2 J^2
\lambda_l^2$ restricting each eigenvalue $\lambda_l^2, l=1,\ldots,n$ of the
averaged squared local susceptibility
$N^{-1}\sum_i\underline{\chi}_{ii}^2$.

To average over the random configurations we have to assume that
equilibrium states determined by the thermodynamic parameters from free
energy~\eqref{eq:FE-final} are non-degenerate. That is, solutions of the
stationarity equations do not depend in the thermodynamic limit on boundary
or initial conditions. This condition is essentially equivalent to
thermodynamic homogeneity. Further on, we assume equivalence of individual
spin components, that is $H^a = H$ and $\langle (S_i^a)^2\rangle_T =
s_n^2$. We then can separate diagonal and of-diagonal matrix elements of
the local susceptibility. Due to the fluctuation-dissipation theorem we can
represent $\chi^{aa} = (s_n)^2 - N^{-1}\sum_i (M_i^a)^2$. We further
introduce the internal magnetic field as an independent random variable
\begin{multline}\label{eq:eta-def}
  \eta_i^a = \sum_j J_{ij} M_j^a - M_i^a\sum_j \beta J_{ij}^2\left ( 1 -
    (M_j^a)^2\right) \\ - \beta J^2\sum_{b\neq a}\chi^{ab} M_i^b \ .
\end{multline}

We add $\sum_i \vec{M}_i\cdot\vec{\eta}_i$ to free
energy~\eqref{eq:FE-final} and subtract the same term with $\eta_i^a$
represented by Eq.~\eqref{eq:eta-def}. We obtain a new representation for
the isotropic case
\begin{multline}\label{eq:FE-isotropic}
  F = \frac N4 \beta J^2 \sum_{a\neq b} (\chi^{ab})^2 + \sum_i \sum_{a=1}^n
  M_i^a\bigg[ \eta_i^a \\ + \frac {\beta J^2}2 \sum_{b\neq a}\chi^{ab}
  M_i^b \bigg] - \frac 14 \sum_{ij}\sum_{a=1}^n \left[2 J_{ij}M_i^a M_j^a
  \right. \\ \left.  + \beta J_{ij}^2 \left (s_n^2 -
      (M_i^a)^2\right)\left(s_n^2 - (M_j^a)^2\right)\right]\\ - \frac
  1\beta \sum_i \ln \left[ \exp\left\{ \frac {J^2}2 \sum_{a\neq b}
      \chi^{ab}\ \frac{\delta}{\delta \eta_i^a}\ \frac{\delta}{\delta
        \eta_i^b}\right\}\right. \\ \left.
    \text{Tr}_S\exp\left\{\sum_{a=1}^n S_i^a(\beta H +
      \beta\eta_i^a)\right\}\right]
\end{multline}
where apart from $M_i^a$ and $\chi^{ab}$ also the fluctuating field
$\eta_i^a$ are variational parameters.

One can prove with the assumptions mentioned at the beginning of this
subsection that the internal magnetic fields $\eta_i^a$ are Gaussian random
variables with covariance
\begin{equation}\label{eq:Fields-gaussian}
\langle\eta_i^a\eta_j^b\rangle_{av} = \delta_{ij}  \sum_l J_{il}^2
M_l^a M_l^b = J^2 q^{ab}\delta_{ij}\ .
\end{equation}

The local Gaussian fields $\eta_i^a$ are used to replace averaging over the
nonlocal spin exchange $J_{ij}$ To make the mean-field free energy explicit
we have to simplify nonlocal terms explicitly dependent on the spin
exchange $J_{ij}$ (only linearly) and terms with local random variables
that are not manifestly Gaussian. Due to the Gaussian character of both the
internal magnetic fields $\eta_i^a$ and the the spin exchange $J_{ij}$ we
obtain
\begin{equation}\label{eq:M-eta}
\langle M_i^a\eta_i^a\rangle_{av} = \sum_b \left\langle \frac {\delta
M_i^a}{\delta \eta_i^b} \right\rangle_{av}
\langle\eta_i^b\eta_i^a\rangle_{av}
\end{equation}
and
\begin{multline}\label{eq:J-MM}
  \langle J_{ij} M_i^a M_j^a\rangle_{av} \\ = \frac {J^2}N \left[
    \left\langle M_i^a \frac{\delta M_j^a}{\delta J_{ij}}
    \right\rangle_{av} + \left\langle \frac{\delta M_i^a}{\delta J_{ij}}
      M_j^a \right\rangle_{av}\right]\ .
\end{multline}
Further on, the nonlocal averaging decays into a product of local averages,
since
\begin{align}\label{eq:Derivatives}
  \frac {\delta M_i^a}{\delta \eta_i^b} = \beta \chi_{ii}^{ab} &,\qquad
  \frac{\delta M_i^a}{\delta J_{ij}} = \beta \sum_b \chi_{ii}^{ab} M_j^b\ .
\end{align}
With the above formulas we can average the free energy over random
configuration so that only averaged site-independent order parameters
remain. The averaged free energy is a sum of local terms only and we can go
over to a free-energy density that, for $n$ component isotropic spin-glass
models, reads
\begin{multline}\label{eq:FE-averaged}
  n f_n = \frac {\beta J^2}4 \sum_{a\neq b} \chi^{ab} (\chi^{ab} + 2
  q^{ab}) - \frac {\beta J^2}4 \sum_{a=1}^n \left((s_n)^2 - q^{aa}\right)^2
  \\ - \frac 1{\beta \sqrt{\det{\underline{q}}}} \int D \eta
  \exp\left\{-\frac 12 \vec{\eta}\cdot
    (\underline{q}^{-1})\vec{\eta}\right\}\\ \ln \left[ \text{Tr}_S
    \exp\left\{ \frac {\beta^2 J^2 }2\sum_{a\neq b}\chi^{ab} S^aS^b +
      \beta\sum_{a=1}^n S^a(H + \eta^a)\right\} \right]\ .
\end{multline}
We denoted $\int D\eta = \prod_{a=1}^n\int_{-\infty}^\infty
d\eta^a/\sqrt{2\pi}$.

The averaged free-energy density~\eqref{eq:FE-averaged} is our starting
point for the analysis of the role of two types of averaged order
parameters: the overlap susceptibilities $\chi^{ab}$ and the overlap
magnetizations $q^{ab}$.  They are determined from saddle-point
(stationarity) equations
\begin{subequations}
\begin{align}\label{eq:overlap-parameters}
  q^{ab}& = \left\langle \text{Tr}_S\left[\rho_\chi S^a\right]
    \text{Tr}_S\left[\rho_\chi S^b\right]\right\rangle_{\eta}\\ \chi^{ab}
  &= \left\langle \text{Tr}_S\left[\rho_\chi S^a
      S^b\right]\right\rangle_{\eta} - q^{ab},
\end{align}
\end{subequations}
where we denoted $\langle\ldots\rangle_{\eta}$ averaging over the random
internal magnetic field $\eta$ and $\rho_\chi$ is the appropriate density
matrix for thermal averaging over spin configurations. We remind that we
assumed equivalence of individual spin components, that is $q^{aa}=q_0$ and
$\chi^{aa} = s_n^2 - q_0$ for $a=1,\ldots,n$.

In general $n$-component spin models, both parameters $q^{ab}$ and
$\chi^{a\neq b}$ can become relevant in the low-temperature phase. To
demonstrate this explicitly we resort our further analysis only to the
replicated Ising spin glass with equivalent replicas, although there are no
principal obstacles to extend the reasoning to more complex models.

\section{Two real replicas}
\label{sec:Finite}

Free energy~\eqref{eq:FE-averaged} can be explicitly evaluated only if we
fix the number of real replicas or components of the vector model. The
simplest situation is the SK model with two equivalent real replicas, that
is, the replicated spin space is isotropic.  In this case we have $(S^a)^2
= 1$, $q^{aa} = q_0$, $q^{a\neq b} = q$, and $\chi^{a\neq b}= \chi$.  It is
then easy to calculate the trace over the replicated spin variables. We
arrive at
\begin{multline}\label{eq:FE-2RR}
  f_2= -\frac {\beta}4 (1-q_0)^2 + \frac{\beta \chi}4(2 q + \chi)\\ -\frac
  1{2\beta} \int \mathcal{D}\xi_1 \mathcal{D}\xi_2 \ln 2 \left\{e_+c_ + +
    e_-c_-\right\}\ .
\end{multline}
We set $J=1$ in this expression and use this energy scale throughout the
rest of the paper. Further on, we used an abbreviation for the Gaussian
differential $\mathcal{D}\phi \equiv {\rm d}\phi\ 
e^{-\phi^2/2}/\sqrt{2\pi}$ and we denoted $e_\pm= \exp\{\pm\beta^2\chi\}$,
\\ $ c_+=\cosh\left[\beta\left(2h + \sqrt{q_0- q^2/q_0}\ \xi_1 +
    (\sqrt{q_0} + q/\sqrt{q_0})\xi_2\right)\right] $, \\ $
c_-=\cosh\left[\beta\left(\sqrt{q_0- q^2/q_0}\ \xi_1 - (\sqrt{q_0} -
    q/\sqrt{q_0})\xi_2\right)\right] $.

With the above notation we can write down the stationarity conditions
representing the equations for the averaged mean-field order parameters.
\begin{subequations}\label{eq:2RR-equations}
\begin{align}
  q_0&= \left\langle\frac{e_+^2s_+^2 + e_-^2s_-^2} {(e_+c_+ + e_-c_-)^2}
  \right\rangle\\
  q &= \left\langle\frac{e_+^2s_+^2 - e_-^2s_-^2} {(e_+c_+ + e_-c_-)^2}
  \right\rangle\\
  \chi &= (e_+^2 - e_-^2) \left\langle\frac 1 {(e_+c_+ + e_-c_-)^2}
  \right\rangle
\end{align}
\end{subequations}
These equations have a solution $q_0 = q$ and $\chi=0$ in the
high-temperature phase. In the low-temperature (spin-glass) phase if
\begin{subequations}\label{eq:2RR-instability}
\begin{align}\label{eq:2RR-instability-chi}
  0 &>1 - \beta^2 \langle (1 - t_{12}^2)(1 - t_2^2)\rangle
\end{align}
then $\chi > 0$ and
\begin{align}
\label{eq:2RR-instability-Delta} 0 &>1 - 4\beta^2 \left\langle
  \frac {(1 - t_2^2)^2} { [1 - t_2^2 + e_-^2(1 + t_2^2)]^2} \right\rangle
\end{align}\end{subequations}
leads to $\Delta = q_0 - q > 0$. We denoted $t_2 = \tanh\beta(h +
\xi_2\sqrt{q_0})$ and $t_{12} = \tanh\beta(h + \xi_1\sqrt{\Delta(2q_0 -
  \Delta)/q_0} + \xi_2(\sqrt{q_0} - \Delta/\sqrt{q_0})$. Hence, either of
these parameters can become an order parameter in the spin-glass phase. We
have to find which one is more relevant for thermodynamic stability.

Numerical analysis and an expansion around the critical temperature show
that we cannot comply with both conditions~\eqref{eq:2RR-instability}
simultaneously. That is, either $\Delta > 0$ and the condition in
Eq.~\eqref{eq:2RR-instability-chi} does not hold ($\chi = 0$), or $\chi >
0$ and then inequality~\eqref{eq:2RR-instability-Delta} cannot be satisfied
($\Delta = 0$). If $\chi = 0$, the free energy does not depend on $\Delta$
and we recover the SK solution with $q =0$ and $q_0 = q_{SK}$.  Hence the
only way to improve upon the SK solution is to choose $\chi> 0$ and $q_0 =
q$. The free-energy density reduces in this case to
\begin{multline}\label{eq:FE-2R}
  f_{2R} = -\frac {\beta}4 (1-q)^2 + \frac{\beta \chi}4(2 q + \chi)\\ 
  -\frac 1{2\beta} \int \mathcal{D}\xi \ln 2\left[ e_+\cosh(2\beta (h +
    \xi\sqrt{q})) + e_-\right]\ .
\end{multline}

\begin{figure}
  \includegraphics[width=8cm]{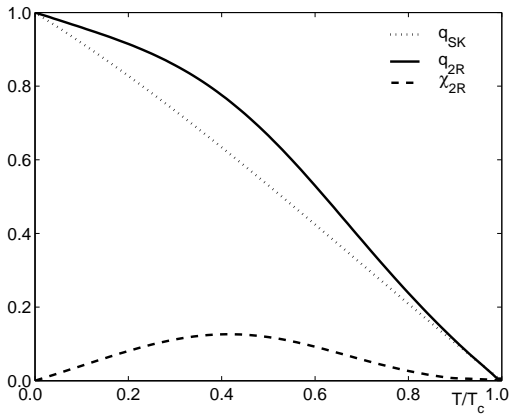}
  \caption{\label{fig:2R-order}Order parameters for the SK model at zero
    magnetic field with two real replicas compared to the SK parameter.}
\end{figure}

\begin{figure}
  \includegraphics[width=8cm] {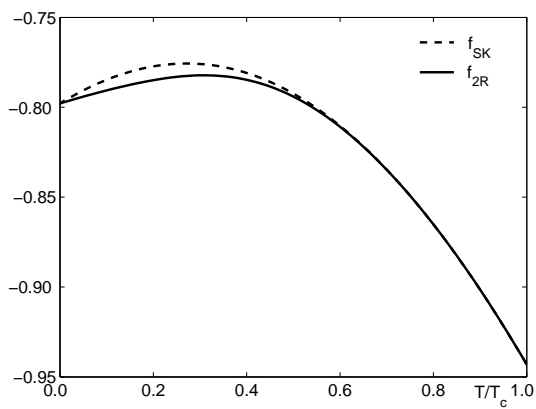} \caption{\label{fig:2R-FE}Free
    energy of the SK model with two real replicas compared to the SK one.}
\end{figure}

It is a straightforward task to solve stationarity equations to
functional~\eqref{eq:FE-2R}. The temperature dependence of the order
parameters is plotted in Fig.~\ref{fig:2R-order} and of the free-energy
density in Fig.~\ref{fig:2R-FE}.

Thermodynamics of the model with two replicas indicates some improvement
upon the SK solution. The order parameter $q = \langle m_i^2\rangle_{av}$
is closer to the Monte-Carlo result.\cite{Kirkpatrick78} Entropy at low
temperatures is seen from Fig.~\ref{fig:2R-FE} to be less negative than
that of the SK solution. However, the free energy is lower than the SK one,
unlike Monte Carlo simulations.  At zero-temperature the free-energy
density has asymptotics $f_{2R} \doteq -\sqrt{2/\pi} + T/4\pi$. It means
that there is no change in the ground-state energy with respect to the SK
result. The negative low-temperature entropy of the SK solution is halved
with two replicas.
 
\begin{figure}
  \includegraphics[width=8cm] {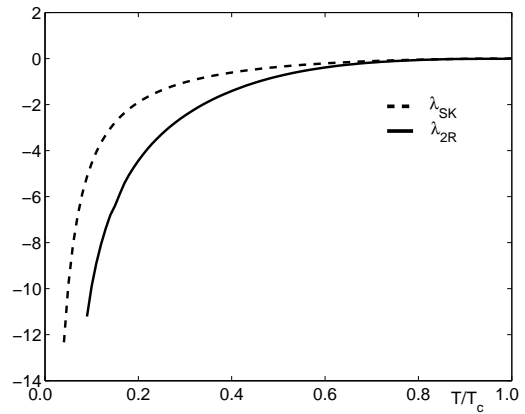}
\caption{\label{fig:2R-instability}Instability  of the two-replica and SK
  solutions in the spin-glass phase at zero magnetic field.}
\end{figure}

The replicated solution must be unstable. A measure of instability is the
generalized AT condition~\eqref{eq:AT-generalized}.  Since we have two
replicas, this equation should stand for two conditions.  One condition is
the complement (negation) to inequality~\eqref{eq:2RR-instability-Delta}.
It is satisfied at any temperature. The second stability condition reads
\begin{subequations}\label{eq:2R-stability}
\begin{align}\label{eq:2R-AT}
  \lambda  &= 1 - \beta^2\langle X_2^2(1 - t_2^2)^2\rangle \ge 0\ ,\\[2pt]
\label{eq:2R-magnetization} X_2 &=  2\ \frac{[1 - t_2^2 + e_-^2(1 +
  t_2^2)]}{[1 + t_2^2 + e_-^2(1 - t_2^2)]^2}
\end{align}\end{subequations}
and is, on the other hand, always broken in the spin-glass phase. When
compared with the AT condition of the SK solution,
Fig.~\ref{fig:2R-instability}, we can see that the two-replica solution
deeper in the low-temperature phase is even less stable than the SK one.
The two-replica solution must be modified to improve upon the instability
of the SK averaged free-energy density.

\section{Analytic continuation - 1RSB}
\label{sec:1RSB}

The most straightforward improvement of the two-replica solution is its
extension to an arbitrary number of equivalent real replicas. This can be
done explicitly provided $\Delta = 0$, that is, $q^{ab} = q$. We show at
the end of the next section that the condition for the instability leading
to $\Delta >0$ is never fulfilled for integer numbers of real replicas.  We
hence assume real-replicas with $\Delta = 0$. The free-energy density then
for a general matrix $\chi^{ab}$ reads
\begin{multline}\label{eq:FE-averaged-finite}
  f_n = \frac{\beta }{4} \left[\frac 1 n\sum_{a\neq b} \chi^{ab}
    \left(\chi^{ab} + 2 q\right) - (1 - q)^2\right]\\ -\frac 1{\beta
    n}\!\int\limits_{-\infty}^{\infty}\mathcal{D} \eta \ln \text{Tr}_S
  \exp\left\{\frac{\beta^2}2\sum_{a \neq b} \chi^{ab}S^aS^b \right. \\
  \left. + \beta \sum_{a=1}^n S^a(H^a + \eta\sqrt{q}) \right\}\ .
\end{multline}

To evaluate explicitly this free energy we assume the replica-symmetric
(isotropic) case, $\chi^{ab} = \chi$ for $a\neq b$, and decouple the
interacting spins in Eq.~\eqref{eq:FE-averaged-finite} by a
Hubbard-Stratonovich transformation linearizing the quadratic term in spin
variables in the exponential. The trace over the spins can be performed
explicitly for arbitrary number of replicas and we obtain a free-energy
density with $n$ real replicas
\begin{multline}\label{eq:FE-1RSB}
  f_n(q,\chi) \\= -\frac \beta4(1-q -\chi)^2 + \frac \beta4 n \chi(2 q +
  \chi) -\frac 1{\beta n}
  \int_{-\infty}^{\infty}\!\!\!\mathcal{D}\eta \\
  \times \ln\int_{-\infty}^{\infty}\!\!\!  \mathcal{D}\lambda
  \left\{2\cosh\left[\beta\left(h + \eta\sqrt{q}
        +\lambda\sqrt{\chi}\right)\right]\right\}^{n}
\end{multline}
We can get rid of the $\lambda$-integration from the Hubbard-Stratonovich
transformation for any positive integer $n$. It is easy to verify that free
energy~\eqref{eq:FE-1RSB} for $n = 2$ reduces to Eq.~\eqref{eq:FE-2R}.
However, the integral representation of the free energy as used in
Eq.~\eqref{eq:FE-1RSB} is indispensable for analytic continuation of the
free energy with integer number of real replicas to an arbitrary real $n$.
This continuation is unique if the limit at $|n|\to\infty$ is analytic. The
limit to infinite-many replicas exists with $\chi\to \chi'/n$. It reduces
the $\lambda$-integration to a saddle-point. Free energy analytically
continued to arbitrary number of real replicas is a fundamental tool for
investigating thermodynamic homogeneity of the solution. According to
Eq.~\eqref{eq:av-homogeneity}, thermodynamically homogeneous free energy
should not depend on the replication parameter $n$.
 
We already know that free energy~\eqref{eq:FE-1RSB} depends on the number
of real replicas. The solution cannot hence be thermodynamically
homogeneous whatever $n$ we choose.  We, however, can use the analytically
continued free energy~\eqref{eq:FE-1RSB} to minimize its inhomogeneity.
This minimization is achieved by vanishing of the local dependence of the
replicated free energy on the replication factor $n$. That is, we demand
\begin{equation}\label{eq:Inhomogeneity-minimum}
\frac{\partial f_n}{\partial n} = 0\ .
\end{equation}
This equation determines the optimal choice of $n$ for which the deviation
from thermodynamic homogeneity of the free energy is minimal.

\begin{figure}
  \includegraphics[width=8cm] {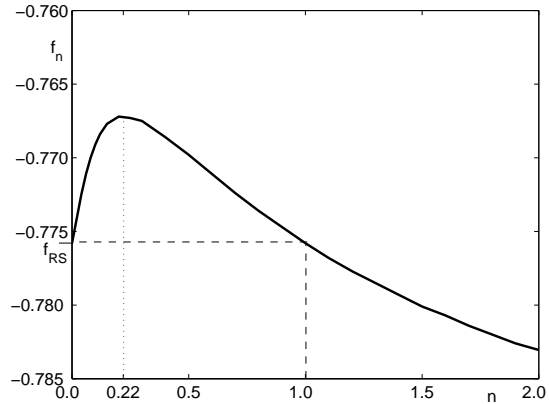}
  \caption{\label{fig:FE-n-dependence} Dependence of the averaged
    free-energy density on the replication parameter $n$ at $T=T_c/4$.}
\end{figure}

Equation \eqref{eq:Inhomogeneity-minimum} has always a solution. It is easy
to verify that $f_0 = f_1 = f_{SK}$, whereby for $n=1$ the free energy does
not depend on $\chi$ and for $n=0$ we have $q_{SK} = q + \chi$. Hence,
there is an $n_{eq}\in [0,1]$ for which
Eq.~\eqref{eq:Inhomogeneity-minimum} is fulfilled. Moreover, since $f_n <
f_{SK}$ for $n > 1$ and $n < 0$, see Fig~\ref{fig:FE-n-dependence}, the
solution with the minimal deviation from thermodynamic homogeneity
maximizes the free energy as a function of the replication factor $n$.

\begin{figure}
  \includegraphics[width=8cm] {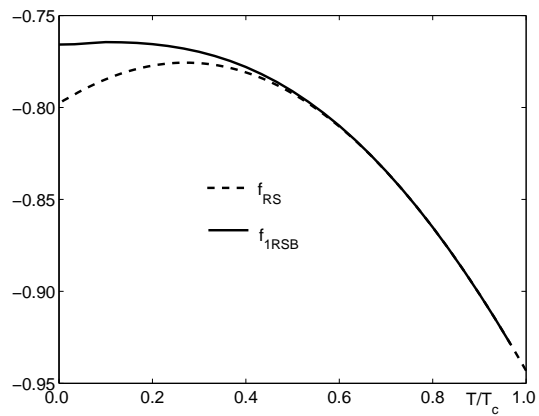}
  \caption{\label{fig:1RSB-FE}Free-energy density for the optimized
    replication parameter compared to the SK solution.}
\end{figure}

\begin{figure}
  \includegraphics[width=8cm] {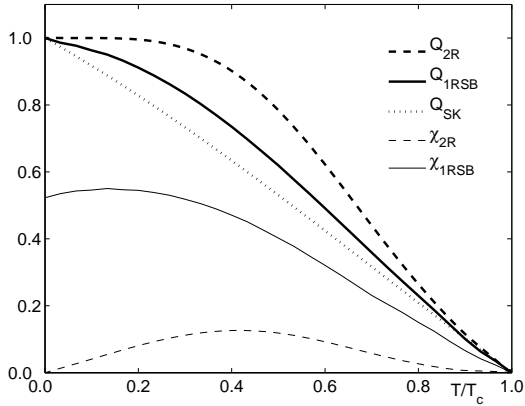}
  \caption{\label{fig:1RSB-order}Order parameters for 1RSB and two real
    replicas. We denoted the Parisi parameter $Q =q + \chi$. The importance
    of optimization of the replication parameter at low temperatures is
    demonstrated on the behavior of the overlap susceptibility~$\chi$.}
\end{figure}

Free energy~\eqref{eq:FE-1RSB} with the optimized replication parameter $n$
via Eq.~\eqref{eq:Inhomogeneity-minimum} is identical with the Parisi 1RSB
solution. It significantly improves upon the SK solution in the spin-glass
phase.  The free energy is higher than the SK one everywhere below the
critical point, Fig.~\ref{fig:1RSB-FE}.
Notice that optimization of the replication parameter $n$ via
Eq.~\eqref{eq:Inhomogeneity-minimum} is a very important constituent of the
RSB solution. In particular in low temperatures. The RSB solution improves
upon the ground-state energy at zero temperature only if the parameter $n$
depends on temperature.  Only if the optimal replication parameter scales
with temperature as $n = T\nu$ we obtain nonvanishing overlap
susceptibility $\chi$ at zero temperature, Fig.~\ref{fig:1RSB-order}, being
responsible for a remarkable improvement in the ground-state energy toward
the Monte-Carlo result.

With $n = T\nu$ we can derive an explicit representation for the
zero-temperature free energy
\begin{equation}\label{eq:FE-zero}
f = - \frac \nu 4 (1-q)^2  - \frac 1\nu \int \mathcal{D} \eta \ln
A(\eta,\nu, q)\
\end{equation}
where we denoted
\begin{multline}\label{eq:A-def}
  A(\eta,\nu, q) \\ = \left[ e^{-\nu\eta\sqrt{q}}
    \text{erfc}\left(-\nu\sqrt{1-q} + \eta\sqrt{q/(1-q)}\right) \right. \\
  \left. +\ e^{\nu\eta\sqrt{q}} \text{erfc}\left(-\nu\sqrt{1-q} -
      \eta\sqrt{q/(1-q)}\right) \right]\ .
\end{multline}
We used the error function $\text{erfc}(x) = \int_x^\infty e^{-t^2/2}
dt/\sqrt{\pi}$. The stationarity solution of the zero-temperature free
energy can be numerically evaluated rather accurately. We obtain $q=0.477
\pm 0.001$, $\nu = 1.351 \pm 0.001$, and $f = -0.76526 \pm 0.00001$.
Theory with a fixed number of real replicas leads at low temperatures for
whatever fixed $n$ (integer or noninteger) to a solution with $\chi =
T\sqrt{2/\pi}/n$, $q = 1 -\chi$, and $f_n = f_{SK} = -\sqrt{2/\pi} + T/2
n\pi \approx -0.798$ as found in the preceding section for $n = 2$.

\section {Stability and hierarchical organization of order parameters}
 \label{sec:Stability}
 
 Although the 1RSB solution improves upon the SK results toward the outcome
 of Monte-Carlo simulations, it is the stability
 condition~\eqref{eq:AT-generalized} that decides whether this could be a
 final form of the mean-field free energy of the SK model. One-step RSB
 solution was derived from Eq.~\eqref{eq:FE-averaged-finite} with an
 isotropic or replica-symmetric ansatz on the overlap susceptibilities. The
 stability condition in this case breaks into two inequalities for two
 eigenvalues of this solution.
 
 A generalization of the AT stability condition~\eqref{eq:2R-stability} for
 arbitrary replication number reads
\begin{subequations}\label{eq:stability-1RSB}
\begin{equation}\label{eq:stability-chi2}
\Lambda_0 = 1  -  \beta^2\left\langle \left[ 1 - (1 -n)  \langle\rho_n
t^2\rangle_\lambda  -   n \langle\rho_n
t\rangle_\lambda^2\right]^2\right\rangle_\eta \ge 0\ .
\end{equation}
We abbreviated $t\equiv \tanh\left[\beta(h + \eta\sqrt{q} +
  \lambda\sqrt{\chi}) \right]$, $\langle X(\lambda) \rangle_{\lambda} =
\int_{-\infty}^{\infty}\mathcal{D}\lambda \ X(\lambda)$ and $\rho_n\equiv
\cosh^n\left[\beta( h + \eta\sqrt{q} + \lambda\sqrt{\chi})
\right]/\left\langle \cosh^n\left[\beta(h + \eta\sqrt{q} +
    \lambda\sqrt{\chi}) \right]\right\rangle_\lambda$.
Inequality~\eqref{eq:stability-chi2} actually stands for the maximal
eigenvalue of the square of the local susceptibility with the
replica-symmetry ansatz. The other eigenvalue is $\langle\langle\rho_n (1 -
t^2)\rangle_\lambda^2\rangle_\eta$.

There is another significant stability condition for the multicomponent
spin-glass theory. It, however, cannot be derived from the TAP approach and
the generalized AT condition~\eqref{eq:AT-generalized}. The second
stability condition is a generalization of
expression~\eqref{eq:2RR-instability-Delta} from two replicas and reads
\begin{equation}\label{eq:stability-Delta}
\Lambda_1 =1- \beta^2\left\langle \left\langle\rho_n \left(1 -
t^2\right)^2 \right\rangle_\lambda\right\rangle_\eta \ge 0\ .
\end{equation}
\end{subequations}
It expresses stability of the replica-symmetry ansatz on the matrix of the
overlap susceptibilities. Should the replica-symmetric solution be stable,
both conditions~\eqref{eq:stability-1RSB} must be obeyed.

\begin{figure}
  \includegraphics[width=8cm] {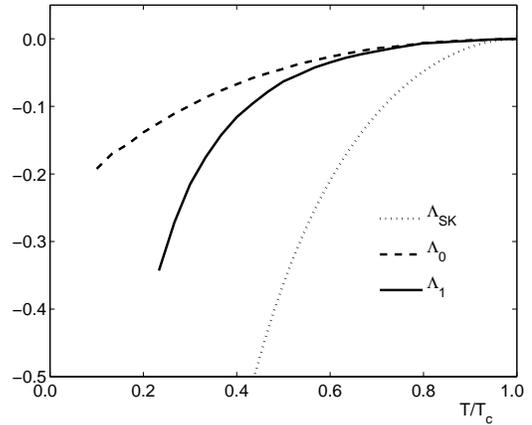}
\caption{\label{fig:1RSB-instability}Instability parameters from the 1RSB
  solution compared with the SK one.}
\end{figure}

To assess the stability conditions quantitatively, we first evaluate the
low-temperature solution near the critical point. There we expand in
$\theta = 1 -T/T_c$ with $T_c = 1$. The 1RSB solution reduces to $q=
\theta/3$, $\chi = 2\theta/3$, $n = 4\theta/3$, and $q_{EA}= q + \chi =
\theta +25\theta^2/27$. Both equations~\eqref{eq:stability-1RSB} show the
same $\theta\to 0$ asymptotics: $\Lambda_0= \Lambda_1 = - 4\theta^2/27$.
On the other hand, the SK solution in this asymptotic limit reduces to
$\langle m_i^2\rangle_{av}= \theta + \theta^2/3$, $\langle m_i^4\rangle = 3
\theta^2$. The AT instability then is $\lambda = 1 - (1 - 2\langle
m_i^2\rangle_{av} + \langle m_i^4\rangle_{av})/(1 - \theta)^2 = -4
\theta^2/3$. The instability of the 1RSB solution is one order smaller than
the instability of the SK one, but nevertheless it is not yet the final
mean-field theory. The full temperature dependence of the the two
instability conditions is plotted in Fig.~\ref{fig:1RSB-instability}.
We can see an overall significant improvement upon the AT instability of
the SK solution in the spin-glass phase.

\begin{figure}
  \includegraphics[width=8cm] {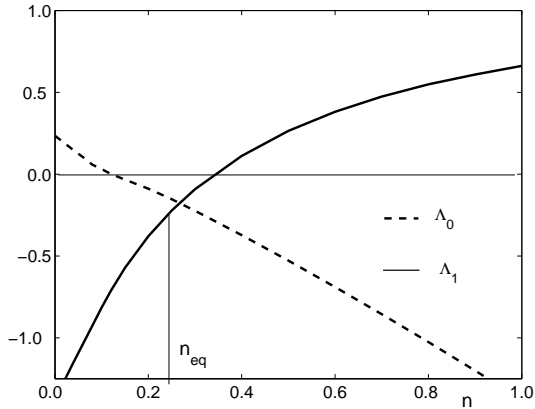}
\caption{\label{fig:1RSB-n-instability}Dependence of the stability
  parameters $\Lambda_0$ and $\Lambda_1$ on the replication parameter $n$
  at $T= T_c/4$. The optimized (equilibrium) parameter $n_{eq}$ for this
  temperature is marked.}
\end{figure}
 
The 1RSB solution, that in this formulation is a replica-symmetric form of
free energy~\eqref{eq:FE-averaged-finite}, is unstable and hence the
isotropic ansatz on the matrix $\chi^{ab}$ was not quite accurate. With the
replica-symmetric ansatz we improved upon the instability of the SK
solution but have not yet reached the global thermodynamic homogeneity of
the averaged free energy.  To find a stable mean-field theory satisfying
the generalized AT condition~\eqref{eq:AT-generalized} we should go beyond
the isotropic ansatz on the matrix of the overlap susceptibilities
$\chi^{ab}$. We can try to make another, less apparent ansatz. It is not,
however, a simple way, since there are several restrictions we have to
comply with. First, there is no other solution except for the replica
symmetric one for $n >1$. Hence, one has to analytically continue the
averaged free energy to the number of replicas less than one. Second, the
number of replicas must be a dynamical variable dependent on the
thermodynamic input parameters, temperature and external magnetic field.
Otherwise we would not improve upon the ground-state energy of the SK
solution at zero temperature. Parisi proposed an ansatz in the effort to
maximize the free energy. To avoid ansatzes, the physical meaning of which
is not quite clear, we can proceed as outlined in
Ref.~\onlinecite{Janis03}. The fundamental idea of this construction is a
hierarchically applied replication of the phase space with the replica
symmetric ansatz on the matrix of the overlap local susceptibilities at
each step (replication). Replication of the phase-space variables actually
checks whether the solution is thermodynamically homogeneous.  Since the
1RSB substantially improved upon the SK solution, we can expect a rapid
convergence toward a globally thermodynamically homogeneous solution in
this way.

We first replicate the spin variables in Eq.~\eqref{eq:FE-averaged-finite}.
We do it so by replacing $n\to n_1 n_2$ and testing homogeneity of the free
energy $f_{n_1}$ with respect to the $n_2$-times enlarged (replicated)
phase space. With the new scaling we have to replicate each spin variable
$S^a$ to $(S^\alpha)^a$ and transform the matrix of overlap
susceptibilities to a super matrix $\chi^{ab}\to(\chi^{\alpha\beta})^{ab}$
where $a,b = 1,\ldots,n_1$ and $\alpha,\beta = 1,\ldots,n_2$. Since all
spin variables $S^a$ are thermodynamically equivalent, they have to split
into new states labeled by the new replica index $\alpha$ identically.

The mean-field equations for the new matrix $(\chi^{\alpha\beta})^{ab}$
contain a replica-symmetric solution $\chi_1 = (\chi^{\alpha\beta})^{ab}$
for $a\neq b$ and $\chi_2 = (\chi^{\alpha\beta})^{aa}$ for $\alpha\neq
\beta$. The superdiagonal matrix element $(\chi^{\alpha\alpha})^{aa}$ is
determined from the fluctuation-dissipation theorem and is not an order
parameter. If the newly won free energy with $q,\chi_1$ and $\chi_2$
physical order parameters and $n_1$ and $n_2$ replication parameters does
not depend on $n_2$ for $q,\chi_1$ and $n_1$ from the 1RSB solution, the
free energy is then globally thermodynamically homogeneous. We, however,
already know that it is not the case for the SK model and the free energy
depends on $n_2$. We again employ the minimization of thermodynamic
inhomogeneity by demanding vanishing of the $n_2$-dependence locally. We
recover the 2RSB solution. We continue in this hierarchical construction so
many times till we reach independence of the free energy of the last
replication parameter. The full RSB solution of Parisi with infinite-many
hierarchical levels for the SK is thereby recovered.

It is worth noting how the stability conditions~\eqref{eq:stability-1RSB}
are related to the generation of new order parameters in the higher level
of the RSB scheme. If the 1RSB solution is unstable, then the new order
parameter $\chi_2$ may emerge in two ways. First, if the factor $\Lambda_0$
from Eq.~\eqref{eq:stability-chi2} is negative and $\Lambda_0 <\Lambda_1$
then $\chi_2$ peels off from zero and $\chi_1 \ge \chi_2>0$.  That is, the
new order parameter $\chi_2$ is the smallest one. If the other parameter
$\Lambda_1 < 0$ has the largest negative value then $\chi_2$ peels off from
$\chi_1$ and $\chi_2 \ge \chi_1>0$.  Note that the overlap susceptibility
$\chi_1$ from the 1RSB solution changes its value in the 2RSB solution with
nonzero $\chi_1$ and $\chi_2$.

The solution is thermodynamically homogeneous only if both
conditions~\eqref{eq:stability-1RSB} are obeyed. We can see in
Fig.~\ref{fig:1RSB-n-instability} that for the equilibrium replication
parameter $n_{eq}$ both conditions are actually broken in the SK model. We
can also see in Fig.~\ref{fig:1RSB-n-instability} that parameter
$\Lambda_1$ is always positive for the replication index $n>1$ and hence
the overlap magnetizations do not split from its diagonal value $q^{ab}=
q^{aa}$ for integer~$n$. Moreover, subsequent replications of solutions
with a fixed (integer) number of real replicas do not lead to real
splitting of the overlap susceptibilities and we have $\chi_1 = \chi_2$ at
the 2RSB level without analytic continuation to $n < 1$.

\section{Interpretation of the replica-dependent order parameters}
\label{sec:Interpretation}
 
An attractive feature on real replicas in the thermodynamic approach of
Thouless, Anderson, and Palmer as used in this paper is a possibility to
avoid a complicated route of averaging over randomness via weighted
contributions from different metastable TAP sates, solutions of the TAP
equations, suggested in Ref.~\onlinecite{DeDominicis83}. Here we treat the
replicated system at low temperatures as if it were in the high temperature
phase. The order parameters, introduced as a response of the system to
infinitesimal symmetry-breaking fields, are hence assumed to determine
uniquely the equilibrium thermodynamic state.  The price we pay for this is
an extension of the phase space and involving the replicated spin variables
in the description of the equilibrium state of the original
(non-replicated) system. To complete the construction of the mean-field
theory with real replicas we have to provide a physical interpretation of
the replica-dependent order parameters.

Having an integer number $n>1$ of replicas of the original spin systems
means that we formally have $n$ copies of the thermodynamic system where
different copies are distinguished by e.g. an external magnetic field. We
assume infinitesimal differences in the magnetic field in different
replicas. The local magnetization from the $a$th replica is $m_i(h^a)$.
The magnetic fields are kept different for finite volumes but acquire the
same value for all replicas after the thermodynamic limit has been
performed. The overlap magnetizations then are $q^{ab} = \lim_{h^a,h^b\to
  h} \langle m_i(h^a) m_i(h^b)\rangle_{av}$. This is the standard way how
real replicas are understood.\cite{Parisi83,Janis87} That is, we assume
that different thermodynamic states with slightly different thermodynamic
parameters cannot be averaged independently.

In this paper we argued that not only configurational averaging cannot be
performed independently for slightly different macroscopic parameters but
also thermal averaging of different spin replicas cannot be performed
separately from each other. Spin values $S_i^b$ from a system replica $b$
influence the local magnetic field in a system $a\neq b$ via the overlap
susceptibility $h^{ab} = \chi^{ab} S_i^b$. Thermodynamic states in one
replica are then influenced by spin configurations in the other system
replicas. Number~$n$ then determines how many different replicas we take
into consideration.

If thermodynamic states in one replica depend on spin configurations in
other replicas, it means that thermodynamic states calculated within one
isolated replica are unstable and are not uniquely determined. To achieve a
stable equilibrium state we have to reach independence of phase-space
replications. Independence of the replica index~$n$ cannot be reached
within the set of integer numbers. Stability conditions for systems with
broken independence of individual thermodynamics in replicated systems
indicate that $n>1$ does not remove the instability of the non-replicated
SK solution.
 
It appears that the only way to improve nontrivially upon the instability
of the non-replicated solution in the real-replica approach is to
analytically continue the replicated theory to noniteger replication
factors $n < 1$.  It cannot, however be done for arbitrary matrices
$q^{ab}$ and $\chi^{ab}$. We hence used the replica-symmetric result from
integer numbers, $q^{ab} = q,\ \chi^{a\neq b} = \chi $. Such an analytic
continuation is then uniquely defined and we can analyze the solution for
arbitrary real $n$.  We find that the averaged free energy density has a
maximum for some equilibrium value $n_{eq}\in [0,1]$. This maximum
corresponds to a locally homogeneous solution, that is, to a solution whose
dependence on the replication index is minimized. This is also the point at
which we maximally improve upon the SK result. The solution with the
optimized $n_{eq}$ coincides with the 1RSB level of the Parisi scheme.

The replication parameter $n < 1$ gives the real replicas another meaning
than do integer values. If $0 < n < 1$, the replication parameter can have
only a probabilistic meaning. We know that the mean-field free energy
depends on the replication number~$n$ if the AT condition is broken and the
TAP equations have many metastable (composite) states, saddle points in the
free energy. They depend on initial/boundary conditions. Since the
macroscopic thermodynamic state is described only by macroscopic
parameters, it must be in a way averaged over all different microscopic
realizations. It means that we observe in thermodynamic equilibrium only
states averaged over the initial conditions.

If solutions of mean-field (TAP) equations depend on initial conditions,
the powers of configurationally dependent variables (local magnetizations)
are not uniquely defined. We have to introduce an additional averaging over
the initial conditions in the TAP approach, i.~e., averaging over spin
configurations from which we started equilibration.  We hence have to
distinguish $\langle m_i^2\rangle_{ini}$ from $\langle m_i\rangle_{ini}^2$,
where the angular brackets denote averaging over the initial conditions. To
perform averaging over the initial conditions properly, we have to weight
the initial spin configurations with the corresponding Boltzmann factor. In
our approach we simulated the weighted averaging over initial conditions by
replicating the spin variables subject to the same thermal equilibration as
the dynamical spins. The replication parameter $0 < n < 1$ then expresses
the probability to find two local magnetizations at the same site with
different values when varying the initial conditions. Such an
interpretation has origin in a decomposition of the local susceptibility
\begin{equation}\label{eq:susceptibility-decomposition}
\chi_{ii} = (1-n)(1 - \langle t_i^2\rangle) + n(1 - \langle t_i
\rangle^2)\ ,
\end{equation}
where $t_i = \tanh\beta(h + \lambda\sqrt{\chi} + \eta_i)$ is the
instantaneous local magnetization and the angular brackets denote averaging
over the internal field $\lambda$, that in our formulation stands for the
weighted averaging over initial conditions.
Definition~\eqref{eq:susceptibility-decomposition} manifests violation of
the fluctuation-dissipation theorem in situations with several possible
values of local magnetizations for TAP macroscopic states. The boundary
value of the replication parameter $n = 0$ corresponds to separated
metastable states (infinite energy barriers between individual states). The
other boundary $n = 1$ leads to a completely random situation with
indistinguishable states (single TAP state). In both limiting cases we
reproduce after averaging the SK solution.
 
Essential part of the construction of a stable solution of mean-field
spin-glass models is minimization of thermodynamic inhomogeneity. The
stable solution should at the end be thermodynamically homogeneous, that
is, independent of the initial conditions. Independence of initial
conditions is marked by independence of the replication factor $n$. If the
stationarity points of the mean-field free energy depend on initial
conditions, thermodynamic homogeneity is achieved by an optimal
organization of different solutions of the mean-field equations. This
organization is reflected in the structure of the matrix of overlap
susceptibilities.  We do not know it \'a priori and have to assume it.

In this paper we discussed only the replica-symmetric choice of the matrix
of the overlap susceptibilities $\chi^{a\neq b} = \chi$. With this ansatz
we assume structureless organization of different solutions of the
mean-field equations.  One can disclose an organization of different
solutions leading to globally homogeneous solution by successive
replications with the replica-symmetric ansatz as proposed in
Ref.~\onlinecite{Janis03}. At present we, however, do not know whether in
this way resulting discrete RSB solution of Parisi is the only
thermodynamically homogeneous macroscopic state.

\section{Conclusions}
\label{sec:Conclusions}

We studied in this paper the genesis of the nontrivial structure of the
order parameters in the RSB solution of the SK model. We derived a
thermodynamic mean-field free energy for typical configurations of the spin
exchange for general vector spin-glass models. In this expression naturally
two types of order parameters emerge. They are the local magnetizations
$m_i^a$ and the averaged local susceptibilities $\chi^{ab}$.  We used this
free energy to analyze a replicated SK model.  We found that due to the
existence of many saddle points in the TAP free energy the local
magnetizations depend on the initial conditions from which we start thermal
equilibration. Due to this fact, powers of the local magnetization are
nontrivially correlated. We concluded from our results that this
correlation is caused by non-negligible dependence of thermodynamic states
(solutions of the TAP equations) on fluctuations in the initial conditions
rather than due to fluctuations in the configurations of the spin-exchange
$J_{ij}$. Hence, the Parisi matrix of order parameters $Q^{ab} = q^{ab} +
\chi^{ab}$ becomes nontrivial because the overlap susceptibility
$\chi^{a\neq b}$ has a complicated internal structure. The disorder-induced
correlation between the local magnetizations $q^{ab} = N^{-1} \sum_i m_i^a
m_i^b = q$ remains structureless in both high- and low-temperature phases.
 
With our analysis we demonstrated that replication of the phase space is a
useful tool for investigating thermodynamic homogeneity of mean-field free
energy for a variety of spin models with or without randomness. Besides
that, when replications are applied successively together with equivalence
of replicas and the replica-symmetric ansatz it offers a systematic
hierarchical construction converging to a globally thermodynamically
homogeneous solution. A substantial ingredient of this construction is
analytic continuation to arbitrary number of real replicas and subsequent
minimization of thermodynamic inhomogeneity if it occurs. If a particular
solution does depend on the replication index, i. e., it is inhomogeneous,
we have to choose such a replication parameter at which the free-energy
dependence on it is minimal.  The optimized free energy is independent of
infinitesimal changes of the replication index. In this way we fully
reconstruct the replica-symmetry breaking solution of the Ising spin glass.

\section*{Acknowledgments}

Research on this problem was carried out within a project AVOZ10100520 of
the Academy of Sciences of the Czech Republic and supported in part by
Grant No. IAA1010307 of the Grant Agency of the Academy of Sciences of the
Czech Republic.

\end{document}